\documentclass[3p,times,procedia]{elsarticle}

\usepackage{ecrc}
\usepackage{siunitx}
\usepackage[font+=scriptsize]{subcaption}

\volume{00}

\firstpage{1}

\journalname{Physics Procedia}

\runauth{}

\jid{phpro}

\jnltitlelogo{Physics Procedia}

\CopyrightLine{2014}{Published by Elsevier Ltd.}

\usepackage{amssymb}

\usepackage[figuresright]{rotating}

\begin{document}

\begin{frontmatter}

\title{Background Assay and Rejection in DRIFT}

\author[csu]{J. Brack}
\author[shef]{E. Daw}
\author[csu]{A. Dorofeev}
\author[shef]{A. Ezeribe}
\author[oxy]{J.-L. Gauvreau}
\author[unm]{M. Gold}
\author[csu]{J. Harton}
\author[unm]{R. Lafler}
\author[unm]{R. Lauer}
\author[unm]{E.R. Lee}
\author[unm]{D. Loomba}
\author[unm]{J. Matthews}
\author[unm]{E.H. Miller\corref{cor1}}
\ead{ehmiller@unm.edu}
\author[oxy]{A. Monte}
\author[edin]{A. Murphy}
\author[boul]{S. Paling}
\author[unm]{N. Phan}
\author[shef]{S. Sadler}
\author[shef]{A. Scarff}
\author[oxy]{D. Snowden-Ifft}
\author[shef]{N. Spooner}
\author[shef]{S. Telfer}
\author[shef]{D. Walker}
\author[csu]{M. Williams}
\author[shef]{L. Yuriev}

\cortext[cor1]{Corresponding author}
\address[csu]{Department of Physics, Colorado State University, CO 80523, USA}
\address[shef]{Department of Physics and Astronomy, University of Sheffield, S3 7RH, UK}
\address[oxy]{Department of Physics, Occidental College, Los Angeles, CA 90041, USA}
\address[unm]{Department of Physics and Astronomy, University of New Mexico, NM 87131, USA}
\address[edin]{SUPA, School of Physics and Astronomy, University of Edinburgh, EH9 3JZ, UK}
\address[boul]{Boulby Underground Science Facility, TS13 4UZ, UK}


\dochead{}


\begin{abstract}
The DRIFT-IId dark matter detector is a m$^3$-scale low-pressure TPC with directional sensitivity to WIMP-induced nuclear recoils.  Its primary
backgrounds were due to alpha decays from contamination on the central cathode.  Efforts to reduce these backgrounds led to replacing the $20$~\si{\micro \meter}
wire central cathode with one constructed from $0.9$~\si{\micro \meter} aluminized mylar, which is almost totally transparent to alpha particles. Detailed modeling of the nature and origin 
of the remaining backgrounds led to an in-situ, ppt-sensitive assay of alpha decay backgrounds from the central cathode. This led to further 
improvements in the thin-film cathode resulting in over 2 orders of magnitude reduction in backgrounds compared to the wire cathode. 
Finally, the addition of O$_2$ to CS$_2$ gas was
found to produce multiple species of electronegative charge carriers, providing a method to determine the absolute position of
nuclear recoils and reject all known remaining backgrounds while retaining a high efficiency for nuclear recoil detection.  \end{abstract}

\begin{keyword}
Radon \sep Dark Matter \sep TPC \sep DRIFT




\end{keyword}

\end{frontmatter}


\section{Introduction}
\label{sec:Intro}

Weakly Interacting Massive Particles (WIMPs) are an attractive dark matter candidate.  
It is thought that these non-baryonic particles form
halos around galaxies including our own.  As the baryonic portion of the galaxy rotates, our solar system passes through this halo at 220~km/s, providing an 
effective wind of WIMPs in the laboratory frame which comes from the constellation Cygnus \citep{Spergel1988}.  

Recoils induced from this WIMP wind exhibit two distinct signatures that can be used to distinguish between a true dark matter signal
and background events.  First, the motion of the Earth around the Sun at 30~km/s causes the apparent speed of the dark matter particles seen from Earth to vary over the 
course of the year and affects the rate of WIMP interactions.  This results in a maximum rate in June and a minimum rate in December; this is the annual modulation.  Second, the rotation of the Earth 
on its axis causes the incoming direction of the dark matter particles to vary over the course of the day in the laboratory frame; this is the sidereal modulation.  
The Directional Recoil Information From Tracks (DRIFT) experiment aims to provide an unambiguous detection of dark matter by observing this signature.  

\section{The DRIFT detector}
\label{sec:DRIFT}

\begin{figure}[ht] 
  \begin{subfigure}[t]{0.47\linewidth}
    \centering
    \includegraphics[width=0.85\linewidth]{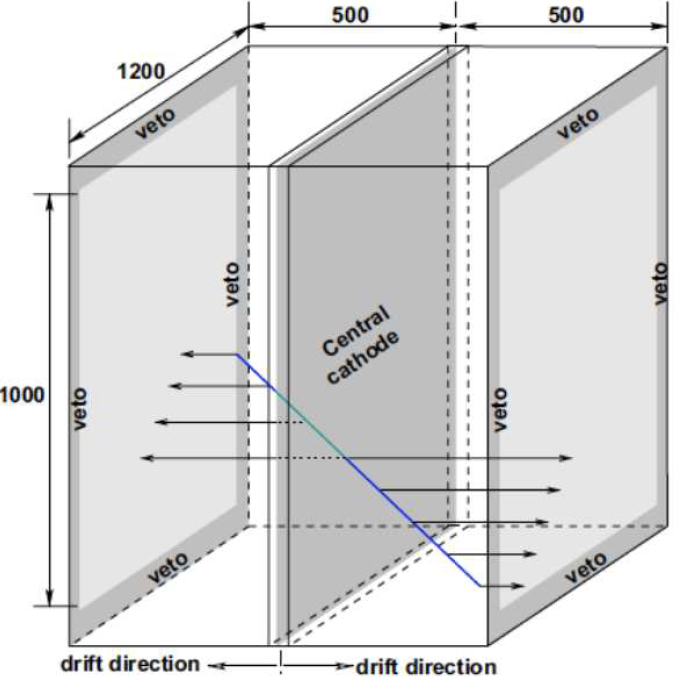} 
    \caption{Schematic of the DRIFT-IId detector, showing a cathode-crossing alpha track.  Lengths are in mm.  Image reproduced from \cite{Burgos2008}.  } 
    \label{fig:DIISchematic} 
  \end{subfigure}
	\hfill
  \begin{subfigure}[t]{0.47\linewidth}
    \centering
    \includegraphics[width=1.1\linewidth]{DIIphoto.png} 
    \caption{Photograph of the DRIFT-IId detector removed from its vacuum vessel.  } 
    \label{fig:DIIPhoto} 
  \end{subfigure} 
  \caption{ The DRIFT-IId detector.}
  \label{fig:DII} 
\end{figure}

The DRIFT-IId detector is a m$^3$ volume low-pressure negative ion gas TPC \citep{Alner2005}.  It uses a mixture of 30 Torr CS$_2$, which offers negative-ion drift to reduce
diffusion \citep{Martoff2000, Snowden2013}, and 10 Torr CF$_4$, which provides spin-dependent WIMP sensitivity via $^{19}$F \citep[e.g.][]{Ellis1991}.  Recently, 1 Torr of O$_2$ has been added as well (see 
Section~\ref{sec:Fiducialization}).  The detector features two back-to-back 50~cm drift volumes which share a central cathode (see Figure~\ref{fig:DII}).  
Each volume ends with a Multi-Wire Proportional Chamber (MWPC) that provides gas amplification and a 3-dimensional readout.  These readouts measure the $x$ and $y$ extent of ionization 
tracks while the outermost wires constitute veto channels which exclude events coming in from the sides.  

The DRIFT-IId detector features excellent range vs. energy discrimination between nuclear recoils, electron recoils, and alpha particles \citep{Snowden2003}.  It is located
1.1~km underground (2800~m.w.e) at the Boulby Underground Science Facility \citep{Murphy2012} to reduce cosmogenic backgrounds, and the vessel is surrounded on all sides with plastic shielding to moderate the 
neutron flux.  Despite these precautions, the detector suffers from inherent backgrounds related to radon, which are described in Section~\ref{sec:Backgrounds}. 
The identification of these backgrounds, and steps taken to reduce them, are described in Sections~\ref{sec:Texturization}-\ref{sec:Fiducialization}.

\section{Backgrounds}
\label{sec:Backgrounds}

DRIFT's dominant backgrounds are due to radioactive contamination undergoing alpha decays on the central cathode \citep{Burgos2007, Daw2010}.  When $^{222}$Rn in the gas decays it produces
a 5.49 MeV alpha particle and a $^{218}$Po daughter.  $22\pm2\%$ of the time the Po daughter is positively charged as it recoils from the decay \citep{Burgos2008}.
This charged atom follows the electric
field to the central cathode where it is electrodeposited.  $^{218}$Po is radioactive and decays, with a half-life of 3.1 minutes, producing another alpha particle and a 
112.2~keV recoiling
$^{214}$Pb atom.  In some cases the alpha particle is oriented toward the central cathode and ranges out in the cathode material.  The $^{214}$Pb  atom, recoiling
in the opposite direction, enters the gas and produces an ionization track very similar to a WIMP-induced nuclear recoil.  This nuclear recoil is called a Radon Progeny Recoil (RPR).  While F and Pb recoils of the same
energy have distinct ranges, events from the central cathode suffer maximum diffusion and the differences between different recoiling nuclei are lost.

The original DRIFT-IId central cathode was made of $20$~\si{\micro \meter} diameter stainless steel wires.  According to geometric calculations with inputs
from SRIM \citep{SRIM}, $36\%$ of the time an alpha particle from a $^{218}$Po decay on the surface 
fully ranges out in a cathode wire and produces an RPR background.  $50\%$ of the time the alpha track is oriented away from the cathode wire,
producing a regular alpha track and embedding the Pb nucleus in the wire.  These events are easily categorized as alpha particles.  The final $14\%$ of decays
produce an RPR in one side of the detector while the alpha particle escapes the wire to produce an ionization track in the other detection volume.  The observation of this alpha track ``tag'' identifies the recoil as coming
from the central cathode, and reveals that it is not due to a WIMP interaction.  

\section{Thin-film cathode}
\label{sec:Texturization}

In March 2010, the DRIFT collaboration replaced the wire central cathode with a $0.9$~\si{\micro \meter} aluminized mylar cathode in order to mitigate the RPRs described in Section~\ref{sec:Backgrounds}.  A typical alpha particle, with a path length of $40$~\si{\micro \meter} in mylar, is fully absorbed only $1\%$ of the time, compared to $30-40\%$ of the time
(depending on the energy of the alpha particle) in a $20$~\si{\micro \meter} stainless steel wire.  This cathode change resulted in a reduction in background rate from $130\pm2$/day to 
only $4.0\pm0.3$/day \citep{Miller2011, Daw2010}.  This change also introduced an unexpected uranium contamination in the new film, described
in Section~\ref{sec:Uranium}.  After a second radiopure cathode was installed in May of 2012 the rate dropped to $2.2\pm0.3$/day.  

\begin{figure}[ht] 
  \begin{subfigure}[t]{0.47\linewidth}
    \centering
    \includegraphics[width=0.95\linewidth]{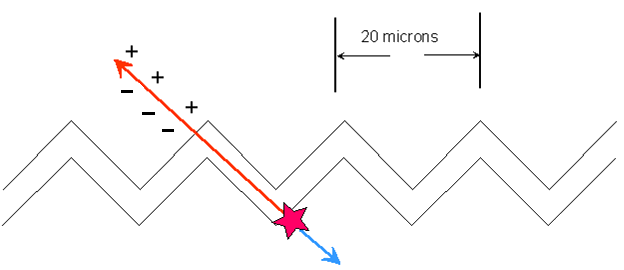} 
    \caption{A thin-film cathode with fine texturization has increased transparency to alpha particles.  } 
    \label{fig:TexMockup} 
  \end{subfigure}
	\hfill
  \begin{subfigure}[t]{0.47\linewidth}
    \centering
    \includegraphics[width=0.95\linewidth]{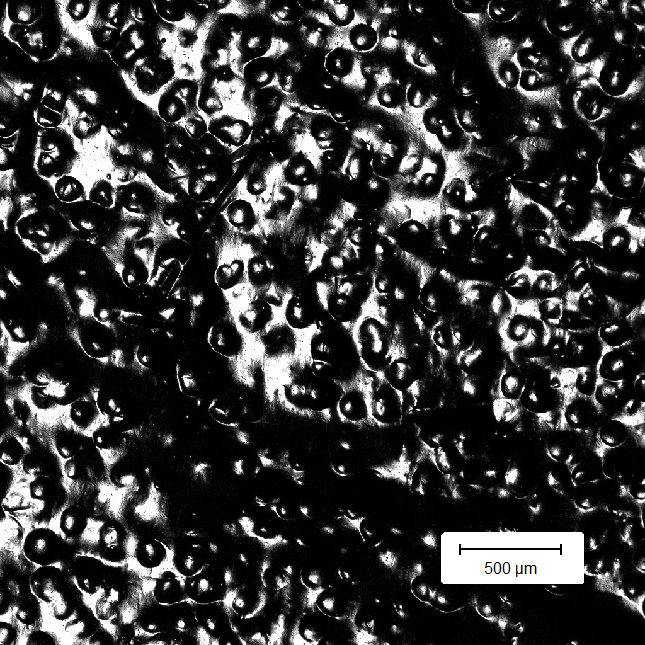} 
    \caption{Confocal microscope image of a texturized thin-film sample similar to the cathode installed in May 2013.   } 
    \label{fig:TexImage} 
  \end{subfigure} 
  \caption{}
  \label{fig:TexFilm} 
\end{figure}

Any further increase in the cathode's transparency to alpha particles would reduce the detector backgrounds.  While an even thinner cathode would
have a higher transparency to alpha particles, mylar thinner than $0.9$~\si{\micro \meter} lacks the strength to be the substrate for a m$^2$-scale cathode.  The
$0.9$~\si{\micro \meter} cathode can, however, be texturized in order to increase its 
transparency to alpha radiation.  The ultimate goal, a $0\%$ chance of
background production, would require complete surface coverage without any straight-line paths longer than $\approx30$~\si{\micro \meter} (see Figure~\ref{fig:TexMockup}).  

The DRIFT collaboration has texturized the thin-film by impacting it with $200-300$~\si{\micro \meter} diameter glass beads \citep{Miller2014}.  Each impact
imparts a nearly hemispherical pit into the mylar.  The longest straight-line path on such a curve that is fully contained within the mylar is only $33$~\si{\micro \meter} long, 
preventing an alpha particle from ranging out in the texturized portion of the cathode.  

The first generation of texturized thin-film cathode was installed in DRIFT-IId in May 2013 (see Figure~\ref{fig:TexImage}), and led to a further reduction in the detector's background
rate from $2.2\pm0.3$~events/day to $0.6\pm0.4$~events/day.  Confocal microscope images of this first generation of texturized thin-film revealed an impact coverage of about $50\%$.  A second generation
of thin-film cathode, more densely texturized than the first, has been manufactured 
for installation in DRIFT-IIe.  DRIFT-IIe is a second underground detector which will begin commissioning in the summer of 2014.

\section{Uranium Contamination}
\label{sec:Uranium}

Section~\ref{sec:Texturization} described the thin-film cathode which was invented to address backgrounds due to alpha decays on the central cathode. 
Nevertheless a detailed picture of the nature of radioactivity at the cathode and how it generates these backgrounds was missing. 
As described in Section~\ref{sec:Backgrounds}, both the radon in the detector and the inherent radioactivity in the cathode material contribute to these backgrounds.  
In order to assess the relative contributions of these, and to better focus efforts towards further reductions in backgrounds, a means for performing in-situ alpha 
spectroscopy was developed. This in turn has led to a further factor of two reduction in backgrounds.

\begin{figure}[ht] 
  \begin{subfigure}[t]{0.47\linewidth}
    \centering
    \includegraphics[width=0.95\linewidth]{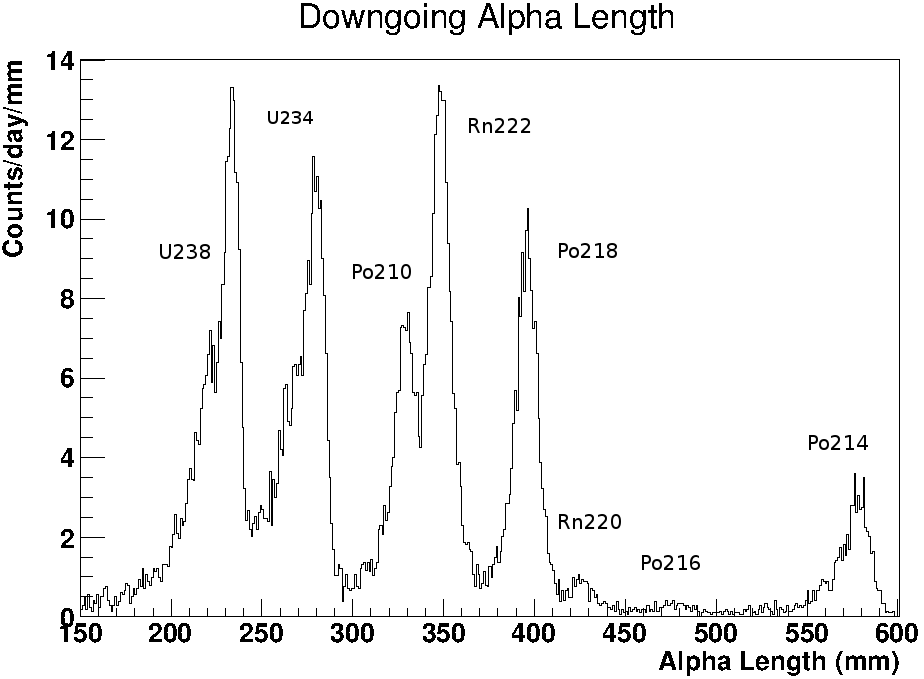} 
    \caption{Spectrum from the first thin-film cathode.} 
    \label{fig:AlphaDownDirty} 
  \end{subfigure}
	\hfill
  \begin{subfigure}[t]{0.47\linewidth}
    \centering
    \includegraphics[width=0.95\linewidth]{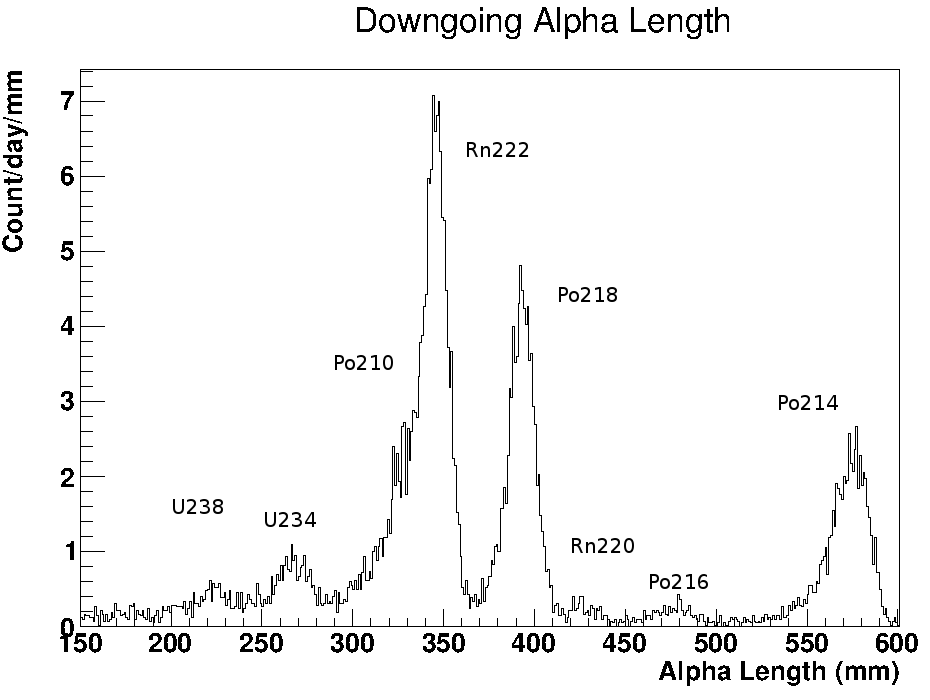} 
    \caption{Spectrum from the second thin-film cathode, demonstrating the reduction in uranium.  } 
    \label{fig:AlphaDownClean} 
  \end{subfigure} 
  \caption{Alpha range spectra from DRIFT allow the identification of radioactive contaminants in the vessel.  Shown are only alpha tracks which are
	down-going (oriented away from the central cathode).  }
  \label{fig:AlphaSpect} 
\end{figure}

The DRIFT detector is designed to measure the lengths of $\approx 50$~keV nuclear recoil tracks, which, in a low-pressure target gas, are typically only 1--2~mm.  
The alpha particles in question, on the other hand, range from 4--7~MeV and have track lengths from 200--600~mm.  DRIFT's range sensitivity, by providing
alpha track length measurements to $2\%$ precision, is better than its energy resolution and allows for range spectroscopy to be used to identify the radioactive isotope.  This
range distribution, with peaks labeled, is shown in Figure~\ref{fig:AlphaDownDirty} and identifies the presence of $^{234}$U and $^{238}$U in the detector
with the first untexturized version of the thin-film cathode deployed.  

DRIFT data can then provide location information for each alpha-producing isotope in a few increasingly precise ways.  
Alpha tracks display a distinct Bragg curve, 
depositing higher ionization density near the head, or end, of the track than at the tail, or origin.  This head-tail measurement indicates if a particular
alpha particle was directed toward the central cathode (``up-going'') or if it was directed toward the MWPC (``down-going'').  These distributions reveal three features: that 
the $^{222}$Rn alpha tracks are equal parts up-going and down-going, as expected for a decay in the bulk gas; that both U populations are exclusively down-going, 
indicating that the contamination is entirely in the central cathode; and that the $^{210}$Po population is mostly up-going, pointing out that the MWPC
is contaminated with this isotope.  

Using the alpha spectroscopy technique, the DRIFT detector can also be used as an in-situ precision assay instrument.
Alpha tracks from decays in the thin-film central cathode, which was designed to be radiation-transparent,
enter the gas $99\%$ of the time.  These events are distinctive high-energy events and are correctly identified as $^{234}$U $35.0\%$ of the time and
as $^{238}$U $33.4\%$ of the time.
The most significant loss of efficiency is due to aspects of the readout which, tailored for short nuclear recoil tracks, is degenerate for alpha tracks oriented
perpendicular to any of the primary axes of the detector.  This reduces the identification efficiency of alpha tracks by about $50\%$, depending on energy.  
In this 21.42 livetime-day exposure, 10,910 $^{234}$U and 6917 $^{238}$U decays were identified, which, accounting for detection efficiencies, corresponds to a 
contamination of the central cathode of $61.8\pm0.6$ ppt and $777\pm15$ ppb, respectively \citep{Miller2014}.  

An alpha decay on the central cathode in which both the alpha track and the RPR are seen in different detector volumes is called a ``tagged'' RPR (see Section~\ref{sec:Backgrounds}).  
While the tag is useful for rejecting the RPR as a potential WIMP interaction, these tagged events also imply that the decay occurred on, or very near to, the surface
of the central cathode; otherwise the recoil would have ranged out in the cathode.  Neither uranium isotope produces significant populations of
tagged RPRs, indicating that these isotopes are located in the bulk of the cathode rather than on its surface.  This alpha spectroscopy technique can provide additional detail on where exactly the uranium contamination lies within the thin-film cathode.

\begin{figure}[ht] 
  \begin{subfigure}[t]{0.47\linewidth}
    \centering
    \includegraphics[width=0.95\linewidth]{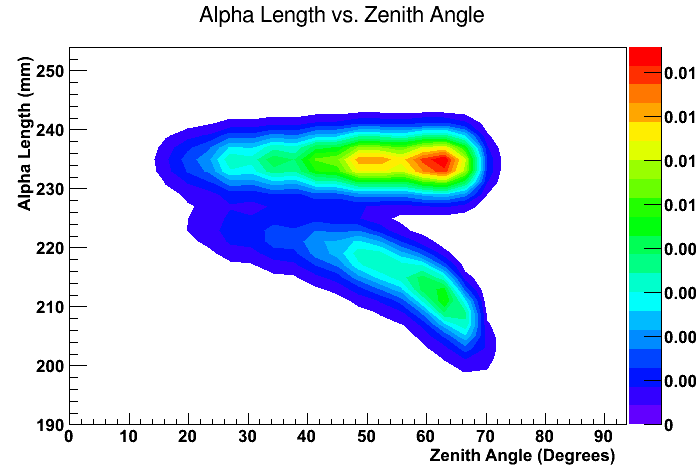} 
    \caption{Uranium in aluminum model.} 
    \label{fig:ModelAl} 
    \vspace{4ex}
  \end{subfigure}
	\hfill
  \begin{subfigure}[t]{0.47\linewidth}
    \centering
    \includegraphics[width=0.95\linewidth]{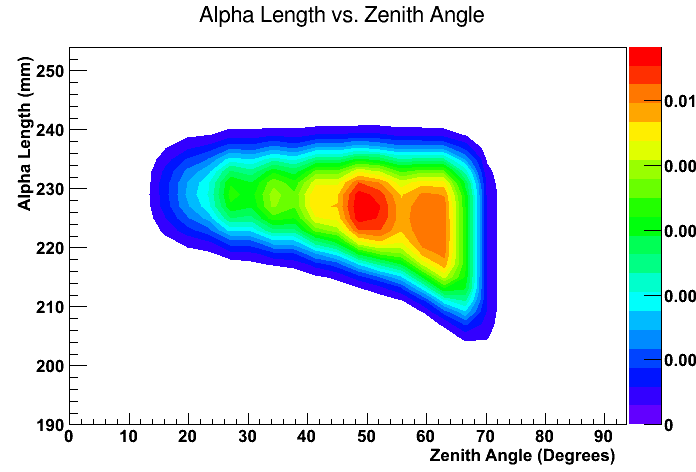} 
    \caption{Uranium in Mylar model.} 
    \label{fig:ModelFilm} 
    \vspace{4ex}
  \end{subfigure} 
  \begin{subfigure}[t]{0.47\linewidth}
    \centering
    \includegraphics[width=0.95\linewidth]{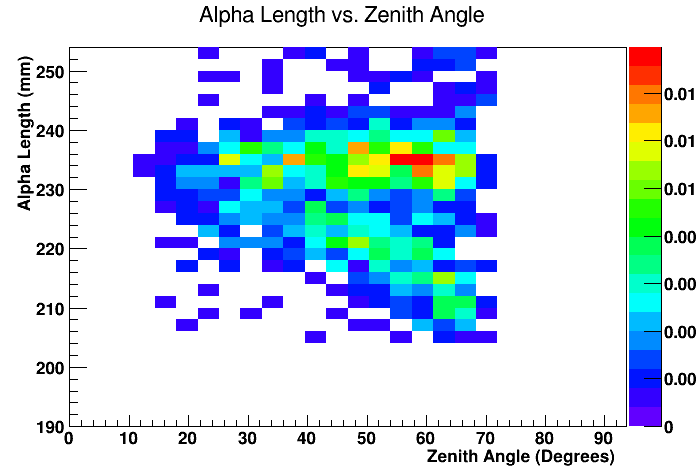} 
    \caption{Data from DRIFT.} 
    \label{fig:Data} 
  \end{subfigure}
	\hfill
  \begin{subfigure}[t]{0.47\linewidth}
    \centering
    \includegraphics[width=0.95\linewidth]{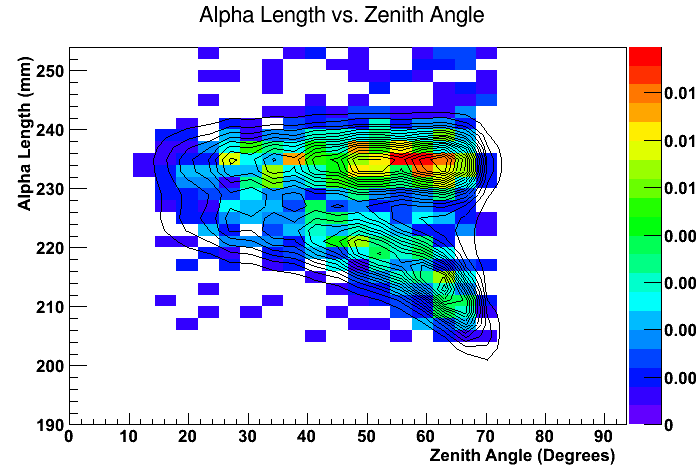} 
    \caption{The data (histogram) overlaid with the best fit model (contour).} 
    \label{fig:ModelOverlay} 
  \end{subfigure} 
  \caption{The shapes of the range vs. angle distribution for uranium indicate that the contamination is in the Al rather than the mylar.  }
  \label{fig:URvTheta} 
\end{figure}

The cathode consists of a $0.9$~\si{\micro \meter} thick mylar substrate with a $0.03-0.05$~\si{\micro \meter} aluminum layer sputtered onto either side to provide electrical conductivity.  DRIFT can distinguish between contamination in the mylar and the aluminum by examining the alpha track length vs.
the alpha track angle $\theta$ measured from the normal to the cathode.  For a population of alpha particles originating from the aluminum layer, this distribution (see Figure~\ref{fig:ModelAl})
has two nodes.  One node corresponds to alpha tracks which exit the cathode on the ``near'' side without passing through the cathode - these particles lose very little
of their energy in the aluminum layer, regardless of angle, so they are all nearly the same length.  The second population of tracks pass through the mylar
and aluminum layers before exiting into the gas on the ``far'' side - these travel through the mylar and lose energy in proportion to $1/\cos(\theta)$.  

If the uranium population is instead distributed evenly throughout the mylar substrate, then the alpha track length vs. $\theta$ distribution will have only one, wider, 
population rather than the above two-node structure (see Figure~\ref{fig:ModelFilm}).  In Figure~\ref{fig:Data} the data plotted are from DRIFT-IId.  These are fit to a linear superposition of the 
two models, which indicates that $90\%$ of the uranium contamination is located in the aluminum coating, while only $10\%$ is from the mylar substrate.  This best-fit is shown along with the data in Figure~\ref{fig:ModelOverlay}.

After determining that the uranium contamination was located predominantly in the aluminum, the DRIFT collaboration constructed a new central cathode using radiopure
aluminum.  Subsequent runs revealed the $^{234}$U ($^{238}$U) contamination reduced by a factor of 20 (10) to $3.3\pm0.1$ ppt ($73\pm2$ ppb) respectively.  After
this switch the background rate also fell from $4.0\pm0.3$/day to $2.2\pm0.3$/day.

\section{$z$-fiducialization}
\label{sec:Fiducialization}

Sections~\ref{sec:Texturization} and~\ref{sec:Uranium} described recent efforts which have reduced DRIFT's backgrounds by over two orders of magnitude.  But the
remaining $\approx1$ event/day background rate would still be a limiting factor for a WIMP search.  These events are understood to all originate at the central cathode, 
and this can be used as a means to reject them if there were a way to measure the event location along the drift axis ($z$).
While the DRIFT-IId detector has always been able to use veto channels on the MWPC to exclude events coming in from the sides of the detector, the third dimension has proven
more difficult.  Until recently, the detector lacked an efficient way of excluding events from the central cathode or MWPCs, as there has been no way to measure the
$z$ position of an individual event.  




\begin{figure}[ht] 
  \begin{subfigure}[t]{0.47\linewidth}
    \centering
    \includegraphics[width=0.95\linewidth]{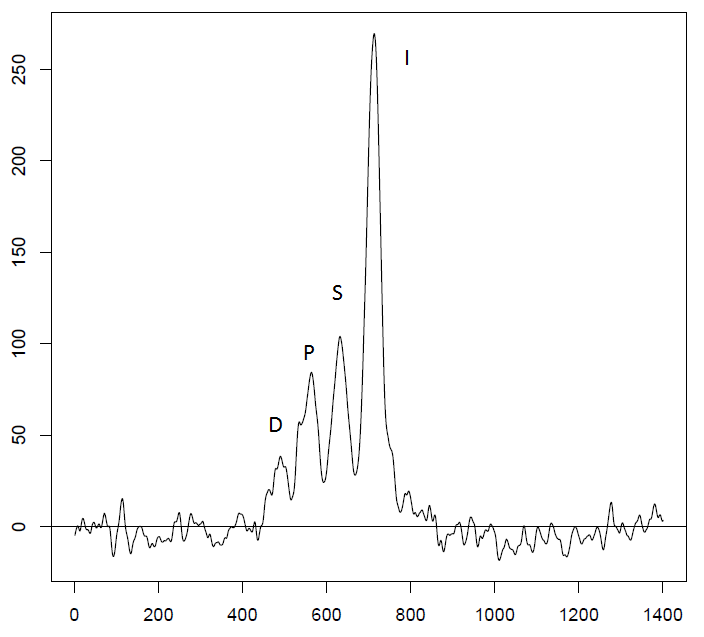} 
    \caption{A neutron-induced nuclear recoil.} 
    \label{fig:RecoilO2} 
  \end{subfigure}
	\hfill
  \begin{subfigure}[t]{0.47\linewidth}
    \centering
    \includegraphics[width=0.95\linewidth]{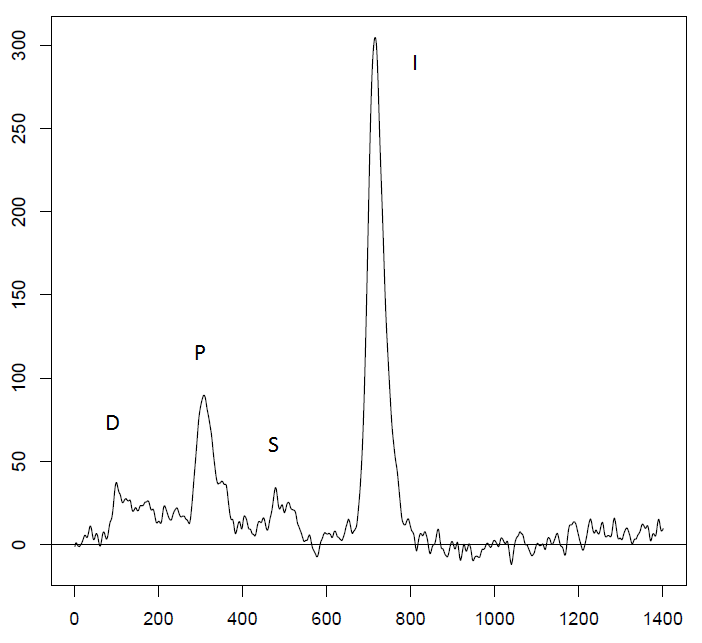} 
    \caption{An RPR from the central cathode.} 
    \label{fig:RPRO2} 
  \end{subfigure} 
  \caption{Two nuclear recoils of similar energy, each with a main peak at 700; times are in units of $\mu$s.  Each waveform has three smaller peaks (S, P, and D) arriving before
	the main peak (I) due to minority carriers \citep{Snowden2014}.  The $z$-location of the event can be determined from the separation of the minority peaks.  }
  \label{fig:RecoilsO2} 
\end{figure}

\begin{figure}[ht] 
  \begin{subfigure}[t]{0.47\linewidth}
    \centering
    \includegraphics[width=0.95\linewidth]{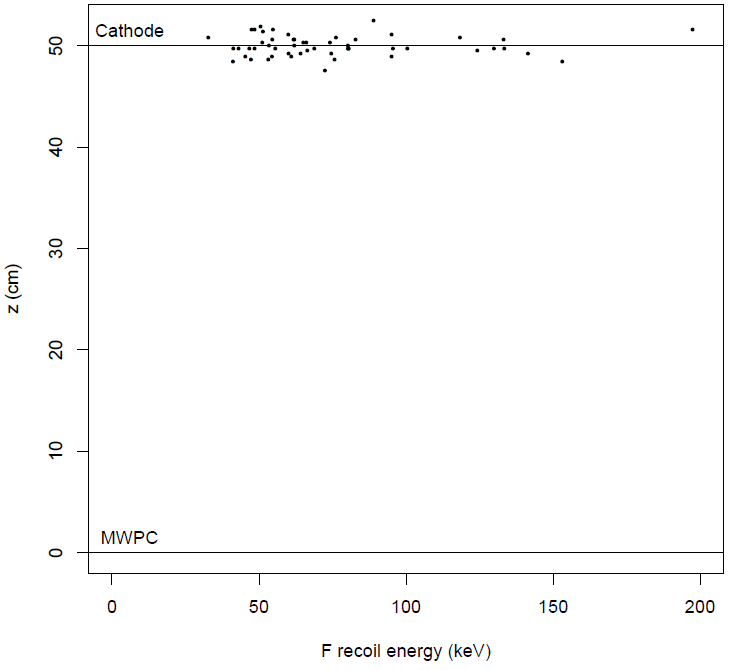} 
    \caption{Events from a WIMP-search run.  All events are located near the central cathode.  } 
    \label{fig:BackgroundO2} 
  \end{subfigure}
	\hfill
  \begin{subfigure}[t]{0.47\linewidth}
    \centering
    \includegraphics[width=0.95\linewidth]{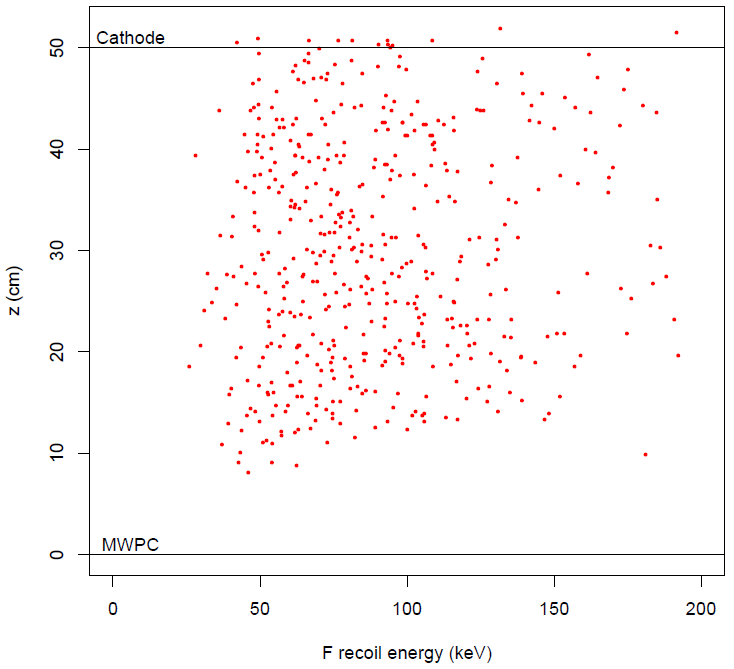} 
    \caption{Events from a neutron exposure.  Interactions are evenly distributed throughout the detector.  } 
    \label{fig:NeutronO2} 
  \end{subfigure} 
  \caption{A comparison of a WIMP-search run and a neutron exposure shows DRIFT's ability to identify all backgrounds as coming from the central cathode while 
  accepting a large fraction of WIMP-like nuclear recoils.  }
  \label{fig:SigRegs} 
\end{figure}

It was recently discovered that trace amounts of O$_2$ mixed with DRIFT's bulk gas, CS$_2$, produces at least three new species of electronegative charge carriers  
whose drift speeds are a few percent faster than the standard CS$_2^-$ ions  (see Figure~\ref{fig:RecoilsO2})\citep{Snowden2014}.  With this additive, the signal on a wire contains four distinct pulses, rather than just one,
and the separation between them increases with the distance that the charge has drifted, thereby providing the $z$ position of the event.  This method to fiducialize
the detector along $z$ provides a powerful means to exclude any background coming from the central cathode.  For events which occur near to the MWPC the peaks
overlap and are difficult to distinguish - in this case the event is rejected, resulting in a several cm dead zone near the MWPC (see Figure~\ref{fig:SigRegs}).  
Figure~\ref{fig:BackgroundO2}
shows the $z$ position vs. energy for events observed during a background run.  For comparison, Figure~\ref{fig:NeutronO2} shows this distribution
for events observed during a neutron exposure.  While the neutrons are distributed throughout the detector, all of the backgrounds are 
clustered around the central cathode and can be removed from any dark matter search with only a modest reduction in WIMP acceptance.

\section{Acknowledgments}
\label{sec:Acknowledgments}
We acknowledge the support of the US National Science Foundation (NSF). This
material is based upon work supported by the NSF under Grant Nos. 1103420 and 1103511. We are grateful to Cleveland Potash Ltd and the Science and Technology
Facilities Council (STFC) for operations support and use of the Boulby Underground
Science Facility.






\bibliographystyle{h-physrev}
\bibliography{TAUPProceedings-full}







\end{document}